\documentclass[epj]{svjour}
\usepackage{graphicx,bm,color}

\newcommand{\sint}{\sin\theta}
\newcommand{\pq}[1]{\left[{#1}\right]}
\newcommand{\p}[1]{\left({#1}\right)}
\newcommand{\ut}{u_\theta}
\newcommand{\up}{u_\phi}
\newcommand{\ulm}{u_{lm}}
\newcommand{\Ylm}{Y_l^m}
\newcommand{\average}[1]{\left<{#1}\right>}
\newcommand{\Hl}{\widetilde H_l}

\newcommand{\rr}{\mathbf{r}}

\newcommand{\derpart}[2]{\frac{\partial #1}{\partial #2}}
\newcommand{\bs}{\bar\sigma}

\begin{document}
\title{On the surface tension of fluctuating quasi-spherical vesicles}

\author{C. Barbetta\inst{1}\thanks{\email{camilla.barbetta@m4x.org}}
\and A. Imparato\inst{2} \and J.-B. Fournier\inst{1}}
\institute{
Laboratoire Mati\`ere et Syst\`emes Complexes (MSC), UMR
7057 CNRS \& Universit\'e Paris Diderot--Paris 7,\\ B\^at.\ Condorcet,
CC 7056, 75205 Paris, France.
\and Department of Physics and Astronomy, 
University of Aarhus, Ny Munkegade, Building 1520,\\
DK-8000 Aarhus C, Denmark}

\date{Received: date / Revised version: date}
%
\abstract{
We calculate the stress tensor for a quasi-spherical vesicle and we thermally average it in order to obtain the actual, mechanical, surface tension $\tau$ of the vesicle. Both closed and poked vesicles are considered. We recover our results for $\tau$ by differentiating the free-energy with respect to the proper projected area. We show that $\tau$ may become negative well before the transition to oblate shapes and that it may reach quite large negative values in the case of small vesicles. This implies that spherical vesicles may have an inner pressure lower than the outer one.}
\maketitle
\section{Introduction}

Biological membranes are nanometer-thick sheets composed essentially of a bilayer of phospholipids~\cite{Mouritsen_book}. Although very thin, they form a barrier for ions and larger molecules. They are thus essential for cells, forming a protection from the external media and preventing diffusion.  Moreover, they are fluid and very flexible, exhibiting a strong resistance to area change but no shear resistance. As a consequence, they present large thermal fluctuations~\cite{Pecreaux04}.  

Bilayers composed of one single type of phospholipid are usually 
used to study the physical properties of membranes. For this purpose, unilamellar vesicles, consisting of one closed bilayer, are widely used in experiments. Different techniques allow to obtain these vesicles, whose size range from a few tens of nanometers to a hundred of micrometers (giant vesicles)~\cite{Hope86,Angelova86,Dimova06}. They appear in several fluctuating shapes, depending on the enclosed volume $V$, on the total area $A$ and on the area difference between the monolayers~\cite{Seifert91,Miao94}. 
 
For liquid interfaces, the surface tension is a constant depending only on the nature of the phases in contact. The situation for membranes is quite different. Membranes consist of a fixed number of insoluble lipids possessing an equilibrium density. Mechanically, the tension depends on the lipid density but also on the curvature of the membrane~\cite{Seifert93,Capovilla02,Fournier07}. Nonetheless, for fluid membranes, the concept of surface tension is threefold: (i) Experimentally, one measures a tension $r$ from the quadratic dependence in the wavenumber of the fluctuation spectrum~\cite{Pecreaux04}. (ii) Theoretically, one can introduce in the Hamiltonian a Lagrange multiplier $\sigma$ multiplying the total membrane area. In principle $r$ can be deduced from $\sigma$ by renormalization procedures~\cite{Peliti85,Cai94}. (More accurately, one may fix the total membrane area by a constraint~\cite{Seifert95} or, even better, keep the lipid density as a parameter~\cite{Seifert93}.) (iii) Experimentally, one measures another tension, $\tau$, through the pressure difference across the membrane by means of the Laplace relation~\cite{Evans90,Fournier08b}. In principle, this is the actual (macroscopic) tension, since it corresponds to the average force that is exerted tangentially to the average membrane's surface. In principle, $\tau$ can be calculated by differentiating the total free-energy with respect to the projected area~\cite{Cai94}, but this is a tricky matter~\cite{Fournier08}. The confusion associated with these different tensions inspired many articles~\cite{Peliti85,Cai94,Fournier08,Kleinert86,David91,%
Fournier01,Henriksen04,Imparato06}.
 
Recently, a new method was introduced to calculate $\tau$, based on thermally averaging the stress tensor of the membrane~\cite{Fournier08}. It has the advantage to be directly related to the very definition of $\tau$. This method was applied to planar membranes and it was shown that the expression of $\tau$ obtained in this way coincides with that obtained from carefully differentiating the total free-energy~\cite{Fournier08}.

The aim of this paper is to calculate the surface tension $\tau$ of quasi-spherical vesicles by using the same method as in~\cite{Fournier08}, i.e., by averaging the stress tensor. There are several interesting questions that we shall address:
\begin{enumerate}

\item What is the difference between $\tau$ for a vesicle and $\tau$ for a planar membrane? Is there a characteristic radius over which they coincide?

\item How does the volume constraint influence the expression for $\tau$?

\item Can one obtain $\tau$ by differentiating the total free-energy with respect to the projected area? If yes, what does projected area mean in the case of a vesicle?
 
\item Can $\tau$ become negative (in this case the Laplace pressure would be inverted with respect to the normal situation)?

\end{enumerate}

Our paper is organized as follows. In sec.~\ref{stress_tensor} we precisely define the system under study and we work out the membrane stress tensor in spherical coordinates. Then we calculate the various correlation functions involved in the vesicle's shape fluctuations. In sec.~\ref{averages}, we use these correlation functions to calculate $\tau$ by thermally averaging the stress tensor. We discuss the relation between $\tau$ and the membrane excess area. In sec.~\ref{free_energy}, we succeed in recovering $\tau$ by differentiating the free-energy with respect to the vesicle's area and we discuss the influence of the volume constraint. In sec.~\ref{discussion} we summarize our results and we discuss the issue of negative tensions.

\section{Stress tensor and fluctuation spectrum}
\label{stress_tensor}

\begin{figure}
  \begin{center}
    \includegraphics[scale=0.7,angle=0]{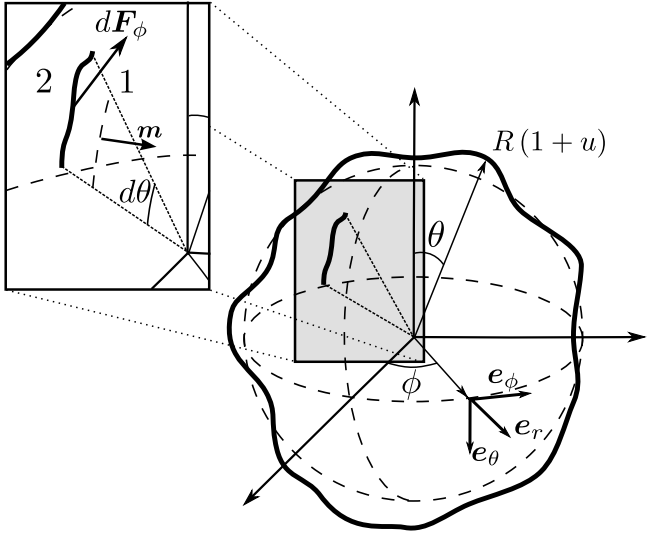}
    \caption{Parametrization in spherical coordinates of a vesicle (bold line) fluctuating around a reference sphere (dashed line). The inset shows the force exchanged through a cut that separates region $1$ and region $2$.}
  \label{param}
  \end{center}      
\end{figure}

\subsection{Effective Hamiltonian and parametrization}

We consider a quasi-spherical vesicle with fixed area $A$ and fixed volume
$V$. Its shape is parametrized in spherical coordinates by (see
Fig.~\ref{param})
\begin{equation}
\bm{r} = R\pq{1 + u\p{\theta,\phi}} \bm{e}_r \,,
\label{param_fluct}
\end{equation}
where $u\ll 1$. A convenient choice for the reference sphere is
$R = (\frac{3}{4}V/\pi)^{1/3}$, corresponding to the
sphere of volume $V$\cite{Seifert95}. 
The volume and area constraints read:
\begin{eqnarray}
V &=& \frac{1}{3}R^3 \int_0^{\pi}\!d\theta\int_0^{2\pi}\!d\phi\,
\pq{1 + u\p{\theta,\phi}}^3\sin\theta \,, 
\label{Volume_0}
\\
A&=&R^2\int_0^\pi\!d\theta\int_0^{2\pi}\!d\phi\left(1+u\right)
\nonumber\\
&&\qquad\times
\sqrt{\left(1+u\right)^2\sin^2\theta+u_\phi^2+u_\theta^2\sin^2\theta}
\,.
\label{Area_0}
\end{eqnarray}
Here, and in the following, $u_i\equiv\partial u/\partial i$, $u_{ij}\equiv\partial^2u/{\partial i\partial j}$, where $i$, $j\in\{\theta,\, \phi\}$. Latin indices will denote either $\theta$ or $\phi$, not $r$.

There are three variants of the free-energy describing the elasticity of a fluid vesicle: i) the ``spontaneous curvature", or Helfrich model~\cite{Helfrich73}, ii) the ``bilayer couple" model~\cite{Evans80,Svetina89}, and iii) the ``area-difference elasticity" model~\cite{Miao94,Wiese92,Bozic92}, the latter being the most accurate~\cite{Miao94}. As shown by Seifert~\cite{Seifert95}, the three models are all equivalent in the quasi-spherical limit to the minimal Helfrich model, i.e., to the spontaneous curvature model with vanishing spontaneous curvature~\cite{Helfrich73}. We therefore adopt the latter, which corresponds to the free-energy, or effective Hamiltonian:
$H=2\kappa\int dS\,C^2$,
supplemented by the area and volume constraints. Here
the integral runs over the vesicle's surface, $C$ is the local mean curvature of the membrane and $\kappa$ is the bending rigidity constant. While the volume constraint is quite easy to implement, it is difficult to handle the surface constraint exactly~\cite{Seifert95}. We shall therefore use the traditional approach, in which the area constraint is approximatively taken into account by means of a Lagrange multiplier $\sigma$ playing the role of an effective surface tension. As discussed in Ref.~\cite{Seifert95}, the latter approach gives correct results in the small excess area limit (in which we shall place ourselves).

Thus, the effective Hamiltonian we shall use is
\begin{equation}
H = \int\!dS \p{2\kappa C^2+\sigma},
\label{helfrich}
\end{equation}
together with the volume constraint~(\ref{Volume_0}). Explicitly, in terms of $u(\theta,\phi)$, the Hamiltonian reads
\begin{equation}
H = \int\!d\theta \,d\phi\, h(u,\{u_i\},\{u_{ij}\}) \, ,
\end{equation}
with~\cite{Helfrich86,Milner87}:
\begin{eqnarray}
h&=&\p{2 \kappa +R^2 \sigma}\sint  \nonumber \\
&+& 2 \sint \left[R^2 \sigma\,u - \kappa \p{u_{\phi\phi}\csc^2 \theta
    +u_\theta\cot\theta +u_{\theta\theta}}\right] \nonumber\\
&+& \frac {1}{2}\sint \left[ 2 R^2 \sigma \,u^2 +(2 \kappa +R^2 \sigma) (\ut^2
  +\up^2 \csc^2 \theta )   \right. \nonumber\\
&+& \kappa \p{\ut \cot\theta +u_{\phi\phi} \csc^2 \theta}^2  +\kappa\,
  u_{\theta\theta}(u_{\theta\theta}+4 u) \nonumber\\ 
&+& \left. 2\kappa (u_{\theta\theta} +2 u)\p{\ut \cot\theta
    + u_{\phi\phi} \csc^2 \theta} \right] + \mathcal{O}(u^3)\, .
\label{energie_2}
\end{eqnarray}

\subsection{Stress tensor components}
\label{tensor_components}

Let us consider an infinitesimal cut at constant longitude ($\phi$ constant) separating a region $1$ from a region $2$ (see Fig.~\ref{param}). The normal to the projection of this cut onto the reference sphere
is $\bm{m} = \bm{e}_\phi$. By definition, the \textit{projected} stress tensor $\bm{\Sigma}$ in spherical geometry relates linearly the force $d\bm{F}$ that region $1$ exerts on region $2$ to the angular length $ds=d\theta$ of the \textit{projection} of the cut onto the reference sphere:
\begin{equation}
d\bm{F}=\bm{\Sigma} \cdot \bm{m} \, ds 
=\p{\Sigma_{\theta\phi}\,\bm{e}_\theta +
\Sigma_{\phi\phi}\,\bm{e}_\phi + \Sigma_{r\phi}\,\bm{e}_r}d\theta.
\end{equation}
Likewise, for a cut at constant latitude ($\theta$ constant),
with $\bm{m} = \bm{e}_\theta$ and $ds=d\phi$, we have
\begin{equation}
d\bm{F}=\bm{\Sigma} \cdot \bm{m} \, ds 
=\p{\Sigma_{\theta\theta}\, \bm{e}_\theta +
\Sigma_{\phi\theta}\, \bm{e}_\phi + \Sigma_{r\theta}\, \bm{e}_r}d\phi.
\end{equation}
For an oblique cut, $d\bm{F}$ is obtained by decomposing $\bm{m}$ along $\bm{e}_\phi$ and $\bm{e}_\theta$.

The projected stress tensor associated with the Hamiltonian of the minimal Helfrich model has been calculated in planar geometry~\cite{Fournier07} and in cylindrical geometry~\cite{Barbetta09} (see also Ref.~\cite{Capovilla02} for the original covariant formulation). The derivation in spherical geometry is lengthier but it follows the same route, which we sketch now. We consider a patch of membrane delimited by a closed curve, corresponding to a domain $\Omega$ on the reference sphere enclosed by $\partial \Omega$. The membrane within the patch is assumed to be deformed, at equilibrium, by means of a distribution of surface and boundary forces (and a distribution of boundary torques). To each point of the patch, we impose an arbitrary displacement $\delta \bm{a} = \delta a_r \, \bm{e}_r + \delta
a_\theta \, \bm{e}_\theta + \delta a_\phi \, \bm{e}_\phi$ that keeps constant along the boundary the orientation of the membrane's normal $\bm{n}$ (in this way the torques will produce no work). On the one hand,
the boundary energy variation reads (after integration by parts):
\begin{equation}
\delta H = \int_{\partial\Omega}\!\! ds\, 
m_i \left[ h \, \delta i + \left(\frac{\partial h}{\partial u_i} -
\partial_i\frac{\partial h}{\partial u_{ij}}\right)\delta u +
\frac{\partial h}{\partial u_{ij}} \, \delta u_j\right],
\label{delta_h_1}
\end{equation} 
where $ds$ is the arc-length in the $(\theta,\phi)$ space, $h$ is given by eq.~(\ref{energie_2}), $\bm{m}$ is the normal to $\partial \Omega$, and $\delta i \in \{\delta \theta,\delta \phi\}$ corresponds to the variation of $\partial\Omega$.
On the other hand, the
work of the force exerted through the boundary reads
\begin{equation}
\delta H = \int_{\partial\Omega}\!\! ds\,\,
 \delta \bm{a} \cdot \bm{\Sigma} \cdot \bm{m}\, .
\label{delta_h_2}
\end{equation} 
By identifying eqs.~(\ref{delta_h_1}) and~(\ref{delta_h_2}), one can obtain
$\bm{\Sigma}$. Details of this calculation, as well as all the components of
the stress tensor are given in appendix \ref{Annexe_1}.
In particular, we obtain 
\begin{eqnarray}
\Sigma_{\theta\theta}&=&
\frac{1}{2R} \Big\{ R^2 \sigma \sint \p{2 + 2 u +
  \up^2 \csc^2 \theta- \ut^2}\nonumber\\
&+&\kappa \left[u_{\phi\phi}^2  \csc^3 \theta - u_{\theta\theta}^2\sint
\right.\nonumber\\
&&\quad+\left. 2\ut\p{u_{\theta \phi\phi}\csc\theta+u_{\theta \theta\theta}\sint}\right] \nonumber \\
&+&\kappa \csc \theta \left[2 \up^2 - \ut^2\cos^2\theta  - 2
    u_{\phi\phi} \p{1+\ut \cot \theta} \right]\nonumber\\  
&+&  4 \kappa\,u\csc \theta \p{u_{\phi\phi} - u_{\theta\theta}\sin^2\theta} \nonumber\\
&+&2 \kappa\pq{  u_{\theta \theta}\p{\sint +\ut\cos \theta}
    + \p{2u-1}\ut\cos\theta} \Big\}\qquad
    \nonumber\\&+&\mathcal{O}(u^3)\,.
\label{stt} 
\end{eqnarray}


\subsection{Thermal averages and fluctuation spectrum}
\label{fluct_spec}

In order to calculate the vesicle's mechanical tension $\tau$, we need to determine the thermal average $\langle\Sigma_{\theta\theta}\rangle$. To this aim, we perform the standard decomposition of $u(\theta,\phi)$ in spherical harmonics~\cite{Seifert95,Helfrich86,Milner87}:
\begin{equation}
u(\theta,\phi) = \frac{u_{0,0}}{\sqrt{4\pi}} 
+ \sum_\omega u_{l,m} Y_l^m(\theta,\phi)\, ,
\end{equation}
with $u_{l, -m} = (-1)^m u_{l,m}^*$ and
\begin{equation}
\sum_\omega \equiv \sum_{l=2}^{L}\sum_{m = -l}^l\,, 
\end{equation}
$L$ being a high wavevector cutoff (see below). Note that the
modes $l=1$, which correspond to simple translations, are discarded.

In terms of $u_{l,m}$ and up to order $u^2$, eq.~(\ref{Volume_0}) takes the form:
\begin{equation}
V = R^3\left[\frac{4\pi}{3}\left( 1 + \frac{u_{0,0}}{\sqrt{4 \pi}}\right)^3 +
  \sum_\omega |u_{l,m}|^2 \right]\, .
\label{Volume}
\end{equation}
The volume constraint $V = \frac{4}{3} \pi R^3$ (recall the definition of $R$) implies therefore~\cite{Seifert95}:
\begin{equation}
u_{0,0} = -\frac{1}{\sqrt{4 \pi}}\sum_\omega |u_{l,m}|^2\, .
\label{u00}
\end{equation}
With the help of the relation:
\begin{equation}
\cot\theta \frac{\partial Y_l^m}{\partial \theta} + \csc^2\theta
\frac{\partial^2 Y_l^m}{\partial \phi^2} = - \frac{\partial^2 Y_l^m}{\partial
  \theta^2} - l\p{l+1} Y_l^m,
\end{equation}
and using eq.~(\ref{u00}), one obtains then~\cite{Seifert95}:
\begin{eqnarray}
H = 4 \pi R^2 \sigma 
+\frac12\sum_\omega\tilde H_l\,|u_{l,m}|^2+\mathcal{O}(u^3)\,,
\label{energie_harm}
\end{eqnarray}
where
\begin{equation}
\tilde{H}_l = \kappa \p{l-1}\p{l+2}\p{l^2+l + \bar{\sigma}}\,.
\end{equation}
Here,
\begin{equation}
\bar{\sigma} = \frac{\sigma}{\kappa/R^2}
\label{sigmabardef}
\end{equation}
is the reduced tension. Note that we have discarded in $H$ a constant energy term, $8\pi\kappa$.

We emphasize that \textit{negative} values of $\bar\sigma$ are allowed~\cite{Seifert95}. Indeed, the minimum of the Hamiltonian~(\ref{energie_harm}) corresponds for $\bar\sigma>-6$ to a perfectly spherical vesicle ($u_{l,m}=0$, $\forall l\ge2$). The mean-field transition to an oblate shape occurs thus at $\bar\sigma=-6$ (non harmonic terms being then needed to stabilize the system).

Standard statistical mechanics yields then $\langle u_{l,m} \rangle = 0$, $\forall l\neq0$, and
\begin{equation}
\langle u_{l,m}\,u_{l',m'} \rangle = (-1)^m \frac{k_\mathrm{B} T}{\tilde{H}_l}\,
\delta_{l,l'}\,\delta_{m,-m'}\,,
\label{correlation}
\end{equation}
where $k_\mathrm{B}T\equiv1/\beta$ is the temperature in energy units.

We may now calculate the fluctuation amplitudes. 
Using eq.~(\ref{correlation}) and the addition theorem for spherical harmonics:
\begin{equation}
\sum_{m=-l}^l Y_l^m(\theta, \phi)\,{Y_l^m}(\theta, \phi)^* = \frac{2l +
  1}{4\pi}\, ,
\end{equation}
we obtain
\begin{eqnarray}
\label{u}
\langle u \rangle &=& \frac{\langle u_{0,0} \rangle}{\sqrt{4\pi}} = -
\frac{1}{4\pi} \sum_\omega \langle |u_{l,m}|^2 \rangle \nonumber\\
&=&-\frac{k_{\mathrm{B}} T}{4 \pi} \sum_{l=2}^L \frac{2l+1}{\tilde{H}_l}\, \\
\label{ucarre}
\langle u^2 \rangle &=& \sum_\omega \sum_{\omega'} Y_l^m(\theta, \phi)
Y_{l'}^{m'}(\theta, \phi) \langle u_{l,m}\,u_{l',m'} \rangle \nonumber \\ 
&=& \sum_{\omega} \frac{k_{\mathrm{B}}T}{\tilde{H}_l} Y_l^m(\theta,\phi)\,
{Y_l^m}(\theta,\phi)^*\nonumber \\
&=& \frac{k_{\mathrm{B}} T}{4 \pi} \sum_{l = 2}^L \frac{2l + 1}{\tilde{H}_l} =
- \langle u \rangle\,
, \\
\langle u_\phi^2 \rangle &=& \sin^2 \theta \, \frac{k_{\mathrm{B}} T}{4 \pi}
  \sum_{l=2}^L \frac{l\p{l+1}\p{2l+1}}{2 \tilde{H}_l}\, ,\\
\langle u_\theta^2 \rangle &=& \frac{k_{\mathrm{B}} T}{4 \pi}
  \sum_{l=2}^L \frac{l\p{l+1}\p{2l+1}}{2 \tilde{H}_l}\,,~\mathrm{etc}.
\end{eqnarray}
The correlations of the other derivatives of $u$ are given in appendix \ref{Annexe_2}.
Note the interesting relation $\langle u\rangle=-\langle u^2\rangle$ (valid at second order in $u$), showing how the temperature-dependent fluctuations affect the mean-shape.

\subsection{Cutoff}

The large wavenumber cutoff $L$ should be related with the smallest wavevector allowed, $\Lambda\approx a^{-1}$, where $a$ is a length comparable to the membrane thickness (i.e., $\pi/\Lambda$ of the order of a few times $a$). With spherical harmonics, however, this is not easy to implement. The requirement that we should recover the planar limit for large values of $R$ will guide us.

For a square patch of fluctuating membrane with reference area $A_p$ and periodic boundary conditions, the wave-vec\-tors are quantified according to $\bm{q} = 2 \pi \sqrt{A_p}\,(n_x,n_y)$, where $n_x$ and $n_y$ are integers and $|\bm{q}| < \Lambda$. The number of modes is then $\pi\Lambda^2/(2\pi/\sqrt{A_p})^2$ and the number of modes per unit area is
\begin{equation}
\frac{N_\mathrm{modes}}{A_p} = \frac{\Lambda^2}{4 \pi}\,.
\label{Nmodes}
\end{equation}
For the vesicle, we have
\begin{equation}
\frac{N_\mathrm{modes}}{A_p} = \frac{1}{4\pi R^2}\sum_{l=2}^L\,\p{2l+1} =
\frac{\p{L-1}\p{L+3}}{4\pi R^2}.
\end{equation}
Asking that the number of degree of freedom per unit area (per lipid, in some sense) be the same in both case, we require these two quantities to be equal. Hence, we get
\begin{equation}
\p{L-1}\p{L+3} = \Lambda^2 R^2\,,
\label{L}
\end{equation}
which gives $L = \lfloor \sqrt{4 + R^2 \Lambda^2} - 1\rfloor$ ($\lfloor x
\rfloor$ is the integer part of $x$). In the limit $R\gg\Lambda^{-1}$, this gives simply $L\simeq \Lambda R$.

\subsection{Validity of the Gaussian approximation}
\label{valid}

Since our calculations are limited to $\mathcal{O}(u^2)$, we should check, in principle, that higher order terms are negligible. In practice this is not feasible. To check the smallness of $u$ (which is especially critical in the case $\sigma\le0$) we shall require:
\begin{equation}
\langle u^2 \rangle = \frac{k_{\mathrm{B}}T}{4 \pi \kappa}\sum_{l=2}^L
\frac{2l +1}{\p{l-1}\p{l+2}\p{l^2 + l +\bar{\sigma}}} \leq U_\mathrm{max}^2\, .
\label{cond_sig}
\end{equation}
To be definite, in the following we shall take
\begin{equation}
U_\mathrm{max}=5\,\%\,.
\end{equation}
Note that the presence of the factor $(l^2+l+\bar\sigma)$ in the denominator of eq.~(\ref{cond_sig}), together with the condition $l\ge2$, implies $\bar\sigma\in[-6,\infty]$, as already discussed.

Solving condition~(\ref{cond_sig}) for the typical values $\Lambda^{-1}\simeq5\,\mathrm{nm}$, $\kappa = 25 \, k_{\mathrm{B}}T$, and taking $U_\mathrm{max}=0.05$, we find
\begin{equation}
\label{val}
\bar\sigma\ge\bar{\sigma}_\mathrm{min}\approx-4\,,
\end{equation}
almost independently of $R$. It follows that negative tensions $\sigma$ are in fact within the validity range of our Gaussian approximation.  

\begin{figure}
  \begin{center}
    \includegraphics[scale=0.75,angle=0]{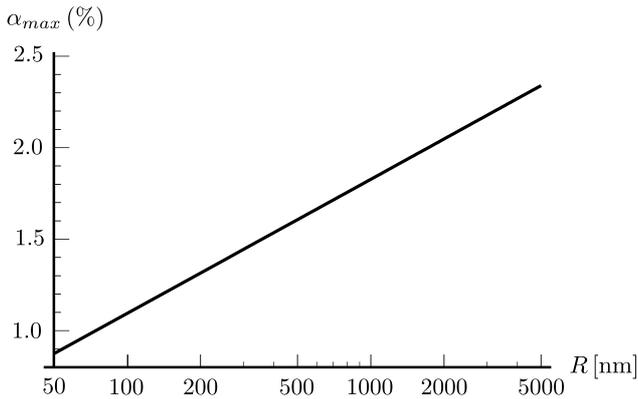}
    \caption{Maximum excess area corresponding to $\sqrt{\langle u^2\rangle}<0.05$, which we take as the validity criterion of our Gaussian approximation. In abscissa is the vesicle's radius. Here, $\Lambda^{-1}\simeq5\,\mathrm{nm}$ and $\kappa = 25 \, k_{\mathrm{B}}T$.}
  \label{alpha}
  \end{center}      
\end{figure}

An interesting control parameter is the excess area, defined by
\begin{equation}
\alpha = \frac{\langle A\rangle - A_p}{A_p} = \frac{k_{\mathrm{B}}T}{8\pi \kappa}
\sum_{l=2}^L \frac{2l+1}{l^2 + l + \bar{\sigma}}\,.
\label{def_alpha}
\end{equation}
Our validity condition~(\ref{val}) yields $\alpha\le\alpha_\mathrm{max}$, with $\alpha_\mathrm{max}$ shown in fig.~\ref{alpha}.
One can see that $\alpha_\mathrm{max} \approx c_1 + c_2\ln R$, where $c_1$ and $c_2$ are constants. Indeed, the sum in eq.~(\ref{def_alpha}) is dominated by the
modes $l=2$ and $l=3$, the rest being well
approximated for $\bar\sigma=\mathcal{O}(1)$ by an integral proportional to $\ln R$.

\section{Actual mechanical surface tension}
\label{averages}

Our aim is to calculate the mechanical tension $\tau$ of the vesicle. Imagine replacing the fluctuating vesicle by a shell coinciding with its average shape. The actual mechanical tension $\tau$ is the average force per unit length that is exerted tangentially to the shell's surface. Because of the spherical symmetry, we may evaluate it at any point in any direction. Let us consider at $(\theta,\phi)$ a cut with $\theta$ constant of extension $d\phi$. The component along $\bm{e}_\theta$ of the force exchanged through the cut is on average $\langle \Sigma_{\theta\theta}\,d\phi\rangle$. The length of the cut is on average $\langle R(1+u)\sin\theta\,d\phi\rangle$. Hence,
\begin{equation}
\tau=\frac{\langle\Sigma_{\theta\theta}\rangle}
{R\sin\theta\p{1+\langle u\rangle}}\,.
\label{tau_def}
\end{equation}
Since $\langle u\rangle=-\langle u^2\rangle$, because of the volume constraint, and since $\Sigma_{\theta\theta}=\sigma R\sin\theta+\mathcal{O}(u)$, we obtain equivalently
\begin{equation}
\label{tau_def_good}
\tau=\frac{1}{R\sin\theta}\langle\Sigma_{\theta\theta}\rangle+\sigma\langle u^2\rangle+\mathcal{O}(u^3)\,.
\end{equation}
Using eq.~(\ref{stt}) and the results obtained in sec.~\ref{fluct_spec} and appendix~\ref{Annexe_2} we obtain
\begin{eqnarray}
\tau &=& \sigma - \frac{k_\mathrm{B} T \kappa}{8\pi R^2}
\sum_{l=2}^L \frac{\p{l-1}l\p{l+1}\p{l+2}\p{2l+1}}{\tilde{H}_l}\quad
\label{resgen}
\\
&=& \sigma - \frac{k_\mathrm{B} T}{8\pi R^2}
\sum_{l=2}^L \frac{l\p{l+1}\p{2l+1}}{l^2 + l + \bar{\sigma}} \, .
\label{tau}
\end{eqnarray}
As expected this result is independent of point $(\theta,\phi)$ where the calculation was made.

\begin{figure}
  \begin{center}
    \includegraphics[scale=0.75]{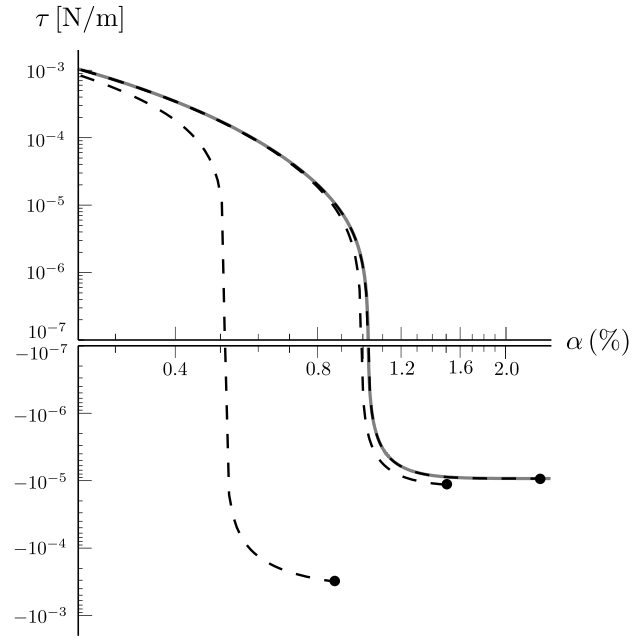}
    \caption{Mechanical tension $\tau$ as a function of the excess area $\alpha$, for $\kappa = 10^{-19}\,\mathrm{J}$, $k_\mathrm{B}T = 4 \times 10^{-21} \, \mathrm{J}$ and $\Lambda = (5\,\mathrm{nm})^{-1}$. The dashed lines correspond from the left to the right to vesicles of radiuses: $R=50\,\mathrm{nm}$, $R = 0.5\,\mathrm{\mu m}$ and $R = 5\,\mathrm{\mu m}$. The solid gray line corresponds to a flat membrane, eq.~(\ref{tauflat}). The end-points indicate the limit beyond which our Gaussian approximation is no longer valid according to sec.~\ref{valid}. The curves for $\tau$ given by eq. (49) (poked vesicles) are qualitatively similar.}
  \label{tau_sigma}
  \end{center}      
\end{figure}

We show in fig.~\ref{tau_sigma} the behavior of $\tau$ as a function of the excess area $\alpha$, obtained numerically from eqs.~(\ref{tau}) and~(\ref{def_alpha}). This presentation is the most physical, because $\tau$ and $\alpha$ are both measurable, while $\sigma$ is not. There are several salient points:
\begin{enumerate}
\item The results for $\tau$ deviate from the flat limit ($R\rightarrow \infty$) essentially for $R\le1\,\mathrm{\mu m}$.
\item Negative and quite large values of $\tau$ are indeed accessible.
\item There exists a well defined excess area corresponding to $\tau=0$. Since, as we shall show, the curves in fig.~(\ref{tau_sigma}) are similar without the volume constraint, this value corresponds to the spontaneous excess area taken up by a poked vesicle (vanishing Laplace pressure).
\item There is a plateau at large values of $\alpha$, which probably corresponds to the actual transition to oblate shapes. It occurs for $\bar\tau\equiv\tau R^2/\kappa<-6$, i.e., below the mean-field threshold (see the discussion after eq.~(\ref{sigmabardef})). The high symmetry phase (spherical vesicle) is thus stabilized by its entropic fluctuations, as one might have expected.
\end{enumerate}

We have also calculated the normal and orthogonal components of the tension. They vanish: $\langle\Sigma_{r\theta}\rangle=0$ and $\langle\Sigma_{\phi\theta}\rangle=0$. While the latter result is obvious on symmetry grounds, the former one is interesting, implying that the shell mentioned above can indeed be considered as a purely tense surface. (This would probably not hold for a vesicle with a non-spherical average shape.) As a consequence, the Laplace law can be used without curvature corrections for a fluctuating quasi-spherical vesicle---if one uses $\tau$ instead of $\sigma$. Indeed, this
could be expected from renormalisation arguments, since the Laplace law is exact (despite the curvature energy) for a perfectly spherical membrane~\cite{Fournier08b}.

It is interesting to examine $\tau$ in the limit of large vesicles. In this
case, the sum in eq.~(\ref{tau}) may be replaced by an integral:
\begin{eqnarray}
\tau - \sigma &\approx& -\frac{k_{\mathrm{B}}T}{8 \pi R^2}
\int_2^{R \Lambda}\!\!dl\,
\frac{l\p{l+1}\p{2l+1}}{l^2 + l + \bar{\sigma}}
\label{tau_int}\\
&\approx& -\frac{k_\mathrm{B}T \, \Lambda^2}{8 \pi}
\left[ 1 - \frac{\sigma}{\kappa \Lambda^2} \ln\left(1 + \frac{\kappa \Lambda^2}{\sigma}\right)\right]
+ \mathcal{O}\left(\frac{\Lambda}{R}\right).
\label{tau_0}\nonumber\\
\end{eqnarray} 
The dominant term in eq.~(\ref{tau_0}) correctly matches the expression of
$\tau - \sigma$ for flat membranes~\cite{Fournier08}. In this case, the relation between $\alpha$ and $\sigma$ is analytical~\cite{Helfrich84}, yielding $\sigma=\kappa\Lambda^2/[\exp(8\pi\beta\kappa\alpha)-1]$ and
\begin{equation}
\tau_\mathrm{flat}(\alpha)=
\frac{\kappa\Lambda^2\p{1+\alpha}}{e^{8\pi\beta\kappa\alpha}-1}
-\frac{k_\mathrm{B}T\Lambda^2}{8\pi}\,,
\label{tauflat}
\end{equation}
which is shown in fig.~\ref{tau_sigma}.

\begin{figure}
  \begin{center}
    \includegraphics[scale=0.65,angle=0]{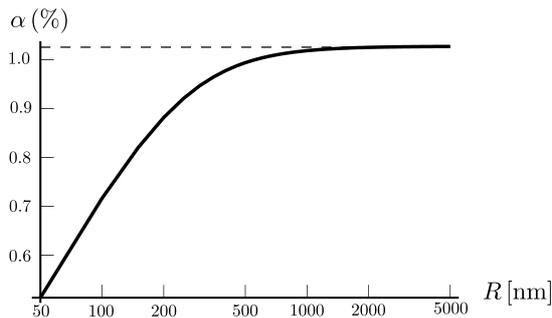}
    \caption{Spontaneous excess area, corresponding to $\tau=0$, as a function of the vesicle radius. The dashed line gives the flat membrane limit $\alpha_0$. The material parameters are the same as in fig.~\ref{tau_sigma}.}
  \label{alpha_spont}
  \end{center}      
\end{figure}

Finally, we show in fig.~\ref{alpha_spont} the excess area corresponding to a vanishing lateral tension ($\tau=0$). Its value is very much radius dependent for $R\le1\mathrm{\mu m}$, but one recovers for $R\ge2\,\mu\mathrm{m}$ the flat membrane limit~\cite{Fournier08}:
\begin{equation}
\alpha_0=\frac{\ln\p{8\pi\beta\kappa}}{8\pi\beta\kappa}\,.
\end{equation}

\section{Derivation using the free-energy}
\label{free_energy}

For a flat membrane, one may also obtain $\tau$ by differentiating
the free-energy with respect to the projected
area~$A_p$~\cite{Cai94,Henriksen04,Imparato06,Fournier08}, but there are two pitfalls~\cite{Fournier08}. One must: i) take the thermodynamic limit ($A_p\to\infty$) only \textit{after} the differentiation, and ii) introduce a variation of the cutoff in order that the total number of modes remains constant during the differentiation, as discussed in~\cite{Imparato06}.

Let us investigate the free-energy method in the case of quasi-spherical vesicles. The free-energy, $F$, is given by
\begin{equation}
F=-\frac1\beta \ln\!\int\!\mathcal{D}[\bm{r}]\, e^{-\beta H},
\label{fe1}
\end{equation}
the integral running over all the configurations of the vesicle.
At the Gaussian level, $H$ is given in terms of spherical harmonics by eq.~(\ref{energie_harm}), and since 
$\bm{r}=R\,\bm{e}_r+R\,u(\theta,\phi)\,\bm{e}_r$, we may write (in agreement with Ref.~\cite{Seifert95}):
\begin{eqnarray}
\mathcal{D}[\bm{r}]
=\prod_{l=2}^L
\p{\prod_{m=0}^l\,R\,du_{l,m}^\mathcal{R}}
\p{\prod_{m=1}^l\,R\,du_{l,m}^\mathcal{I}},
\label{jacob}
\end{eqnarray}
where the superscripts $\mathcal{R}$ and $\mathcal{I}$ signify real part and imaginary part, respectively.
This measure corresponds to the so-called normal gauge, which is known to be correct for small fluctuations~\cite{Seifert95}. 
We note that the radius $R$ of the reference sphere appears explicitly and that for each value of $l$, only half of the allowed values of $m$ have to be considered, as $\rr$ is real.

Performing the Gaussian integrals yields
\begin{equation}
F= 4 \pi R^2 \sigma +k_B T \sum_{l=2}^L\frac{2 l +1}{2}\ln\p{\beta \Hl/R^2} \, .
\label{eq_free}
\end{equation} 
In order to obtain $\tau$, we must differentiate
$F$ with respect to the vesicle's ``projected area" $A_p$. Which one, however? The area $4\pi R^2$ of the reference sphere (i.e., the sphere having the same volume as the vesicle's), or the area  of the vesicle's average shape, defined as $A_m=4\pi\langle R(1+u)\rangle^2$? It will turn out that the former choice is the correct one. In a sense, this is natural because it corresponds to our parametrization. However, it is not that obvious, because the definition of $\tau$ in eq.~(\ref{tau_def}) involves the area of the \textit{average} vesicle's shape.

Let us thus pick $A_p=4\pi R^2$. It is worth noticing
that $\Hl$ depends on $A_p$ only through $\bar{\sigma}=\sigma R^2/\kappa$, yielding
\begin{equation}
\derpart{\Hl}{R^2}=\p{l-1}\p{l+2}\sigma\,.
\end{equation} 
With the choice:
\begin{equation}
\tau=\derpart{F}{A_p}=\frac1{4\pi}\derpart{F}{R^2}\,,
\end{equation}
we obtain
\begin{equation}
\tau
=\sigma- \frac{k_B T}{8 \pi R^2}\sum_{l=2}^L
\frac{l\p{l+1}\p{2l+1}}{l^2+l+\bs}\,,
\end{equation} 
which is identical to the result obtained from the stress tensor approach,
eq.~(\ref{tau}). How about the pitfalls mentioned above? First, we didn't take the thermodynamic limit before differentiating. (Actually, this would not be problematic, since the quantification of the modes does not involve the size of the system, like it is the case for planar membranes.) Second, we kept $L$ (hence the number of modes) constant during the differentiation, in agreement with the fact that $L = \lfloor \sqrt{4 + R^2 \Lambda^2} - 1\rfloor$ is constant for a mathematically infinitesimal change of $R$.

We may obtain a more intrinsic expression for~$\tau$. With $\average{A}=A_p(1+\alpha)$, where $\alpha$ is given by
eq.~(\ref{def_alpha}), and $N_\mathrm{modes} = \sum_{l=2}^L (2l+1)$, we
may rewrite $\tau$ as
\begin{equation}
\tau A_p=\average A \sigma -\frac{k_B T}{2 }N_\mathrm{modes}\,.
\end{equation} 
The quickest way to this result is to keep separate, when differentiating with respect to $R^2$, the two terms coming from $\ln(\tilde H_l)$ and $\ln(1/R^2)$ in eq.~(\ref{eq_free}). The interpretation of this equation is not straightforward, because $\frac12 k_\mathrm{B}T$ is the internal energy per mode (not the free-energy per mode). Note that the same form for $\tau$ is also valid in the planar case, as shown in~\cite{Imparato06} (with further justifications in~\cite{Fournier08}).
 
In addition, we have checked that differentiating $F$ with respect to $A_m=4\pi [R(1+\langle u\rangle)]^2$, which differs from $4\pi R^2$ by terms of $\mathcal{O}(u^2)$, gives a wrong result (in the sense that it differs from the result obtained by the stress tensor method).

\subsection{Poked vesicle}

In the case where the volume is unconstrained, e.g., a vesicle poked by a micropipette or by specific proteins, the natural choice for the reference sphere (of radius $R$) is the average vesicle's shape. By definition, then, $u_{0,0}=0$, which implies $A_m=4\pi R^2$. In this case, the Hamiltonian in the Gaussian approximation becomes:
\begin{eqnarray}
H' = 4 \pi R^2 \sigma 
+\frac12\sum_\omega\tilde H'_l\,|u_{l,m}|^2\,,
\label{energie_harm2}
\end{eqnarray}
where $\tilde H'_l=\tilde H_l+4\kappa\bar\sigma$~\cite{Henriksen04}.
Equation~(\ref{resgen}), which gives $\tau$ by the stress tensor method, is valid whatever the form of $\tilde H_l$. Hence, we need just replace $\tilde H_l$ by $\tilde H_l+4\kappa\bar\sigma$, which yields:
\begin{equation}
\tau'
=\sigma- \frac{k_B T}{8 \pi R^2}\sum_{l=2}^L
\frac{l\p{l+1}\p{2l+1}}{l^2+l+\bs+\frac{4\bs}{\p{l-1}\p{l+2}}}\,.
\label{tau2}
\end{equation}
(To avoid confusions, we denote here by $\tau'$ the mechanical tension in the case of poked vesicles.)
Although eqs.~(\ref{tau2}) and (\ref{tau}) differ mathematically, it turns out that their difference is numerically irrelevant. Indeed the extra term $4\bar\sigma/[(l-1)(l+2)]$ in the denominator of eq.~(\ref{tau2}) is only important for small $l$'s while the sum is dominated by large $l$'s.

The free-energy, is given by the same expression as eq.~(\ref{eq_free}) with $\tilde H_l$ replaced by $\tilde H'_l$: 
\begin{equation}
F' = 4 \pi R^2 \sigma + k_\mathrm{B} T
\sum_l \frac{2l+1}{2} \ln\left(\beta \tilde H'_l/R^2\right)\, .
\label{eq_free2}
\end{equation}
It turns out, again, that $\tau'=\partial F'/\partial (4\pi R^2)$ exactly. 
This result is satisfying, but at the same time it shows how slippery the free-energy approach can be: differentiating with respect to the area of the average vesicle is correct in the case of poked vesicles but not in the case of closed vesicles. The stress tensor method is thus a much safer method.  

Our expression for $\tau'$ differs from that obtained in Ref.~\cite{Henriksen04}, where the authors considered also a spherical membrane without volume constraint. In particular, the mechanical tension obtained in that reference cannot take negative values. We believe that the discrepancy between the two results comes from the omission in Ref.~\cite{Henriksen04} of the factors $R$ within the measure. Indeed the factor $1/R^2$ in the logarithm of our eq.~(\ref{eq_free2}) is absent in the corresponding expression~(A.9) of Ref.~\cite{Henriksen04}.

\section{Summary and discussion}
\label{discussion}

In this paper we have applied the stress-tensor method~\cite{Fournier08} to the calculation of the surface tension $\tau$ of quasi-spherical vesicles, investigating both closed vesicles (volume constraint) and poked vesicles (no volume constraint). In both cases we have shown that the more standard method based on differentiating the free energy with respect to the projected area gives the same results, provided it is performed in the right way.

In order to describe the elasticity of the vesicle we have used the simplest Helfrich model, since all its variants (ADE, BC, etc.) are equivalent in the quasi-spherical limit~\cite{Seifert95}. We have made two approximations: i) we have replaced the membrane area constraint by a Lagrange multiplier approach with effective surface tension $\sigma$; ii) we have kept all calculations at the Gaussian level. The former approximation is justified for small excess areas (see sec.~4C of Ref.~\cite{Seifert95}). To control the validity of the latter approximation, we have restricted our investigations to relative height variations of the membrane with respect to the reference sphere of less than $5\%$.

We have calculated the projected stress tensor in spherical geometry. This tool should be useful in other contexts. We have thermally averaged it in order to calculate the actual surface tension $\tau$ acting tangentially along the average membrane surface. Its expression in terms of $R$ (vesicle radius), $\sigma$, $\kappa$, $k_\mathrm{B}T$, $\Lambda$ (cutoff) depends on whether the vesicle is closed or poked. However, numerically the difference is practically indistinguishable. The flat membrane limit holds when $R\ge1\,\mathrm{\mu m}$. We have re-obtained the same expressions for $\tau$ by the free-energy approach. This confirms that the so-called naive measure that we have used is valid in the Gaussian approximation (see, e.g., Ref.~\cite{Cai94}). We have discussed whether the free-energy should be differentiated with respect to the area of the sphere that has the same volume as the vesicle or with respect to the sphere corresponding to the average vesicle's shape. We have shown that the correct choice depends on whether the vesicle is closed or poked, and therefore that the stress tensor approach is the safest one.

It was shown in Ref.~\cite{Seifert95} that $\sigma$ (the Lagrange multiplier) may become negative. One of our motivation was to check whether the actual mechanical tension $\tau$ could become negative or not. This would imply that vesicles could sustain inner pressures \textit{lower} than the outer pressure (which is impossible for liquid drops). Our analysis shows that $\tau$ may indeed become negative within the validity range of our Gaussian analysis. For instance, small vesicles with $R=50\,\mathrm{nm}$ can reach negatives tensions of $-10^{-4}\,\mathrm{N/m}$ with an excess area still less than $1\%$ (see fig.~\ref{tau_sigma}). The so-called ``giant vesicles", with $R\ge1\mathrm{\mu m}$, can sustain negatives tensions $\simeq\!-k_\mathrm{B}T\Lambda^2/8\pi$ which may be of order $-10^{-6}\,\mathrm{N/m}$ or $-10^{-5}\,\mathrm{N/m}$ depending on the uncertainty on the value of the cutoff. The possibility to sustain negative tensions, or negative pressure differences, might be investigated: i) by controlling the outer osmotic pressure, in the case of small vesicles, or ii) by poking a giant vesicle through by a micropipette to which it would adhere and gently decreasing its inner pressure.

An interesting feature, which might be tested experimentally by controlling the pressure outside the vesicle, is the plateau of fig.~\ref{tau_sigma}. One sees that when $\tau$ reaches a critical value $\tau_c<0$, the excess area rises dramatically. As discussed above, this behavior (still within the validity range of our Gaussian approximation) probably corresponds to the transition to oblate shapes. For small vesicles we find that it corresponds roughly to $\tau_c R^2/\kappa\approx-5$ while for giant vesicles it is given by $\tau_c\simeq-k_\mathrm{B}T\Lambda^2/(8\pi)$. Hence, applying the Laplace pressure formula $\Delta P=2\tau_c/R$, we find that the critical pressure difference yielding the shape transition is
$\Delta P_c\approx-\sup[10\kappa/R^3,k_\mathrm{B}T\Lambda^2/(4\pi R)]$.

Finally, a possible way to measure the difference between $\tau$ and $\sigma$, suggested to us by L. Limozin~\cite{Limozin}, would be to pursue experiments on adhering vesicles similar to those of Ra\"edler et al.~\cite{Raedler95}. In these experiments, the tension is deduced both from the fluctuation spectrum of the adhering part and from the contact angle $\vartheta_\mathrm{eff}$ via the Young-Dupr\'e relation, the adhesion energy being estimated using the distance to the substrate. From our understanding, the former method should give $\sigma$ and the latter one should give~$\tau$. By way of illustration,
the results of Ref.~\cite{Raedler95} for $\sigma\simeq1.7\times10^{-5}\,\mathrm{J}/\mathrm{m}^2$ gave an adhesion energy $\sigma[1-\cos(\vartheta_\mathrm{eff})]\simeq1.3\times10^{-9}\,\mathrm{J}/\mathrm{m}^2$ significantly larger than the value $W_a=8\times10^{-10}\,\mathrm{J}/\mathrm{m}^2$ estimated from the membrane--substrate separation. The authors of Refs~\cite{Raedler95} and \cite{Seifert95b} point out that this problem is recurrent. If the correct relation is rather $\tau[1-\cos(\vartheta_\mathrm{eff})]=W_a$ then we obtain $\tau\simeq1.1\times10^{-5}\,\mathrm{J}/\mathrm{m}^2$, which is compatible with the formula $\sigma-\tau\simeq k_\mathrm{B}T\Lambda^2/(8\pi)$ [see eq.~(\ref{tau_0})] if one takes $\Lambda^{-1}\simeq5\,\mathrm{nm}$. More experimental studies along these lines would be welcome.

\begin{acknowledgement}
\centerline{***}

A.I. gratefully acknowledges the hospitality and support
of the Laboratoire Mati\`ere et Syst\`emes Complexes (MSC) of the Universit\'e
Paris Diderot--Paris 7 where this work was initiated.
\end{acknowledgement}

\appendix

\section{Derivation of the stress tensor}
\label{Annexe_1}

In Sec.~\ref{tensor_components}, we explained the general method to derive the
stress tensor. Here we give the details of this calculation. In order to compare eqs.~(\ref{delta_h_1}) and~(\ref{delta_h_2}), one needs to obtain $\delta\theta$, $\delta\phi$, $\delta u$, $\delta u_\theta$ and $\delta u_\phi$ in terms of $\delta\bm{a}$. The first three can be obtained by identifying the new membrane shape $\tilde{u} = u + \delta u$ with the old one translated: $R[1 + \tilde{u}(\theta + \delta \theta, \phi + \delta\phi)] \, \bm{e}_r(\theta + \delta \theta, \phi + \delta \phi) = R[1 + u(\theta,\phi)]\, \bm{e}_r + \delta \bm{a}$. This leads to 
\begin{eqnarray}
\label{delta_a_r}
\delta a_r &=&
R\p{u_\theta \, \delta \theta + u_\phi \, \delta\phi + \delta u},\\
\delta a_\theta &=& R\p{1 + u} \delta\theta \, ,\\
\delta a_\phi &=& R\p{1 + u} \sin\theta \, \delta \phi\,.
\label{delta_a_phi}
\end{eqnarray}
Let $\delta\bm{n} = \tilde{\bm{n}}(\theta + \delta
\theta, \phi + \delta \phi) - \bm{n}(\theta,\phi)$ be the variation of the
normal, where $\tilde{\bm{n}}$ is the normal to the shape defined by 
$\tilde{u}(\theta,\phi)$ and $\bm{n} = \bm{t}_\theta \times \bm{t}_\phi/|\bm{t}_\theta\times \bm{t}_\phi|$, with $\bm{t}_i = \partial_i \bm{r}$. The variation of the normal vanishes (implying no work of the torques) if $\delta \bm{n}
\cdot \bm{t}_\theta = 0$ and $\delta\bm{n} \cdot \bm{t}_\phi = 0$ over the
border. This leads to
\begin{eqnarray}
\delta u_\theta &=& \frac{1}{1+u}\,\Big\{
\delta \theta \big[\p{1+u}^2+ 2 u_\theta^2-\p{1+u}u_{\theta\theta}\big]
\nonumber\\
&+&\frac{\delta \phi}{\sin\theta}
\big[ \p{1+u}u_\phi\cos\theta-\p{1+u}u_{\theta\phi}\sin\theta 
\nonumber\\
&+& 2u_\phi u_\theta \sin\theta\, \big] + \delta u \,  u_\theta \Big\}\,,
\label{eq4}
\\
\delta u_\phi &=& \frac{1}{1+u}\,\Big\{
\delta\theta\,\big[\p{1+u}u_\phi \cot\theta+ 2 u_\theta u_\phi
\nonumber\\
&-&\p{1+u}u_{\theta\phi}\big] + \delta\phi \big[\p{1+u}^2\sin^2\theta+2 u_\phi^2
\nonumber\\
&-&\p{1+u}\p{u_\theta\cos\theta\sin\theta +u_{\phi\phi}}\big] 
+ \delta u \, u_\phi \Big\}\,.
\label{eq5}
\end{eqnarray}
These equations, combined with eqs.~(\ref{delta_a_r}--\ref{delta_a_phi}),
allow us to write $\delta u_\theta$ and $\delta u_\phi$
in terms of $\delta \bm{a}$. Up to order $u^2$, we obtain
\begin{eqnarray}
\delta\theta &=& \frac{\delta a_\theta}{R} \p{1-u+u^2},\\
\delta\phi &=& \frac{\delta a_\phi}{R\sint }\p{1-u+u^2},\\
\delta u &=& \frac{1}{R} \left[\delta a_r-\p{1-u}\p{\delta a_\phi u_\phi\csc\theta
  +\delta a_\theta  u_\theta}\right],\\
\delta u_\theta &=& \frac{1}{R} \Big\{\delta a_r \p{1-u}u_\theta+\delta
    a_\theta\big[1+u_\theta^2-\p{1-u}u_{\theta\theta}\big]\nonumber\\
&+& \delta a_\phi\csc\theta \big[u_\phi \pq{\p{1-u}\cot\theta
    +u_\theta}-\p{1-u}u_{\theta\phi}\big]\Big\}\,,\nonumber\\\\
\delta u_\phi &=& \frac{1}{R} \Big\{\delta a_\theta\big[u_\phi\pq{\p{1-u}\cot\theta +u_\theta}-\p{1-u}u_{\theta\phi}\big]\nonumber\\
&+& \delta a_r \p{1-u}u_\phi + \delta a_\phi\csc\theta\big[\sin^2\theta +u_\phi^2
\nonumber\\
&-& \p{1 - u}\p{u_{\phi\phi}+u_\theta\cos\theta \sin\theta}\big]\Big\}\,. 
\end{eqnarray} 
These expressions are to be inserted into eq.~(\ref{delta_h_1}). Note that it is necessary to expand $h$ at $\mathcal{O}(u^3)$ in order to obtain
$\partial h/\partial u_i$ and $\partial h/\partial u_{ij}$ consistently at
$\mathcal{O}(u^2)$ in eq.~(\ref{delta_h_1}). This means adding 
\begin{eqnarray}
h_3 &=& - \kappa \sin\theta \left\{ 4 \csc^4\theta\, u_\phi\,u_\theta\,
  u_{\theta \phi} + 2 u_\theta^2\,u_{\theta\theta} \right.\nonumber \\
&-& 2 u_\phi^2 \csc^2\theta\left(u_{\phi\phi}\csc^2\theta - u_\theta\cot\theta
 \right) \nonumber \\
&+& u\left[u_{\theta\theta}^2 + u_\theta^2\left(2 + \cot^2\theta\right) +
  2u_\phi^2\csc^2\theta + u_{\phi\phi}^2\csc^4\theta\right.\nonumber\\
&+& \left. 2 u_\theta\,u_{\theta\theta}\cot\theta 
+ 2 u_\theta\,u_{\phi\phi}\csc^2\theta \cot\theta\, 
+ 2 u_{\theta\theta}\,u_{\phi\phi}\csc^2\theta\right]\nonumber\\
&+& \left. 2 u^2 \left(u_{\theta\theta} + u_\theta\cot\theta 
+ u_{\phi\phi}\csc^2\theta\right)\right\} \, , 
\end{eqnarray} 
to eq.~(\ref{energie_2}) before calculating the derivatives. Comparing eqs.~(\ref{delta_h_1}) and~(\ref{delta_h_2}), we obtain finally
\begin{figure*}
\rule{18cm}{0.4pt}
\begin{eqnarray}
\Sigma_{\theta\theta}&=& \frac{1}{2R} \big\{ R^2 \sigma \sint \p{2 + 2 u +
  \up^2 \csc^2 \theta - \ut^2} +\kappa \pq{u_{\phi\phi}^2  \csc^3 \theta 
  - u_{\theta\theta}^2\sint+ 2 \ut\p{u_{\theta \phi\phi}\csc\theta
  +u_{\theta\theta\theta}\sint}}\nonumber \\
&&\qquad+  \kappa \csc \theta \pq{2 \up^2 - \ut^2\cos^2\theta - 2
    u_{\phi\phi} \p{1+\ut \cot \theta} + 4 u \p{u_{\phi\phi} 
    - u_{\theta\theta}\sin^2\theta}}\nonumber\\
&&\qquad +2 \kappa\pq{  u_{\theta \theta}\p{\sint +\ut\cos\theta}
    + \p{2u-1}\ut \cos\theta} \big\} \, , 
\label{stt_anexo}\\
\Sigma_{\phi\theta} &=& - R \sigma u_\theta u_\phi + \frac{\kappa}{R} \left(2
  u_{\theta \phi} - u_\phi\cot\theta\right) + \frac{\kappa}{R} \left(4
  u\,u_\phi\cot\theta - u_\phi\,u_{\phi\phi}\cot\theta \csc^2\theta + u_\theta
u_\phi - 4 u\,u_{\theta \phi} - u_{\phi\phi}\,u_{\theta\phi}\csc^2\theta
\right.\nonumber\\
&&-\left.u_\theta\,u_{\theta\phi}\cot\theta 
+ u_\phi\,u_{\theta \phi \phi}\csc^2\theta 
+ 2 u_\phi\,u_{\theta\theta}\cot\theta -
  u_{\theta\phi}u_{\theta\theta} + u_\phi u_{\theta\theta\theta}\right) \, ,
\label{spt}\\
\Sigma_{r\theta}&=& R \sigma u_{\theta}\sin\theta -
\frac{\kappa}{R}\left[\p{1-2u}u_\theta\left(2\sin\theta - \csc\theta\right) +
  u_{\theta \theta}\cos\theta - 2 u_{\phi\phi}\cot\theta \csc\theta +
  u_{\theta\phi\phi}\csc\theta + u_{\theta\theta\theta}\sin\theta
 \right] \nonumber\\
&&-\frac{\kappa}{R} \left[ 2u_\phi^2\cot\theta \csc\theta -
  u_\theta^2\cos\theta + 4 u\,u_{\phi \phi}\cot\theta \csc\theta 
  - 2u\,u_{\theta\theta}\cos\theta\right. \nonumber\\
&&\qquad \left. - 2 u\,u_{\theta\phi\phi}\csc\theta 
- 2u\,u_{\theta\theta\theta}\sin\theta 
+ u_{\phi\phi}\,u_\theta\csc\theta- 3 u_\theta\,u_{\theta\theta}\sin\theta\right] \, ,
\label{srt}\\
\Sigma_{\phi\phi}&=& \frac{1}{2R} \left\{ R^2 \sigma \p{2 + 2 u -\up^2 \csc^2
  \theta + \ut^2}+\kappa \pq{-u_{\phi\phi}^2  \csc^4 \theta +
    u_{\theta\theta}^2+ 2 \up\csc^2\theta\p{u_{\phi\phi\phi}\csc^2\theta 
    + u_{\theta \theta\phi}}}\right. \nonumber \\
&&\qquad  + \kappa \csc^2 \theta \pq{2 \up^2+ \ut^2\p{3 \sin^2\theta - 1}+ 2 u_{\phi\phi}\p{1-\ut\cot\theta -2 u}}+ 2 \kappa \p{1-2u}\ut\cot\theta
-2 \kappa \p{1-2u} u_{\theta\theta}\nonumber\\ 
&&\qquad \left. +2\kappa \up\,u_{\theta \phi} \csc^2 \theta \cot\theta
\right\} \,, 
\label{spp} \\
\Sigma_{\theta\phi} &=& - R \sigma u_\theta\,u_\phi\csc\theta + 
\frac{\kappa \csc\theta}{R} 
\left[ \, \p{u_\phi \cot\theta-u_{\theta\phi}}\left(-2 + 4 u +
      u_{\theta\theta} + u_{\phi\phi}\csc^2\theta\right) 
      + u_\theta\,u_\phi\p{1 + \csc^2\theta} \right .\nonumber\\
&+& \left. u_\theta (u_{\phi\phi\phi}\csc^2\theta
  + u_{\theta\theta\phi})\right] \,,\\
\Sigma_{r\phi} &=& \frac{\csc\theta}{R} \left[ u_\phi \left(R^2 \sigma - 2
    \kappa + 4 \kappa u + 3 u_{\phi\phi}\csc^2 \theta + 3u_\theta \cot\theta
    - u_{\theta\theta}\right) - \kappa \p{1 - 2u}\p{u_{\theta\phi}\cot\theta +
  u_{\theta\theta\phi} + u_{\phi\phi\phi}\csc^2\theta}\right] \, .
\label{srp}
\end{eqnarray} 
\rule{18cm}{0.4pt}
\end{figure*}
\begin{eqnarray}
\mbox{\textit{see eqs.~(\ref{stt_anexo}--\ref{srp}).}}
\nonumber
\end{eqnarray}
These expressions are valid up to $\mathcal{O}(u^2)$.

\subsection{Planar limit}

In order to check the expressions~(\ref{stt_anexo}--\ref{srp}), we consider the limit $R\rightarrow\infty$, which should yield the stress tensor for a flat membrane~\cite{Fournier07}. We consider a general point $(\theta, \phi)$ on the sphere and we define a local system of cartesian coordinates $(x,y,z)$ such that $dx=R\,d\theta$, $dy=R\sint\,d\phi$, and $dz=dr$.
We want to determine the stress tensor $\bm{\Sigma}'$ in these new coordinates. The component of the elementary force $df_\alpha$ along a general direction
$\alpha\in\{x,y,z\}$ exerted through a cut perpendicular to $x$ reads $df_\alpha=\Sigma'_{\alpha x}\,d\ell=\Sigma_{\alpha\theta}\,d\phi$, with the correspondence $d\ell=R\sint\,d\phi$. Thus we have $\Sigma'_{\alpha x}= \Sigma_{\alpha \theta}/(R \sint)$. Likewise, $\Sigma'_{\alpha y}= \Sigma_{\alpha \phi}/R$ since in this case $d\ell=R\,d\theta$. In the limit of large $R$, the plane tangent to the sphere at the point $(\phi,\theta)$ becomes the reference plane of the membrane, and the height of the membrane over this plane is $h(x,y)=R\,u(\theta,\phi)$.
Hence, $\ut=R\,u_x=h_x$, $\up=R\,\sint\,u_y=\sint\,h_y$, $u_{\theta\theta}=R^2\, u_{xx}=R\,h_{xx}$, $u_{\phi\phi}=R^2 \sin^2\theta\,u_{yy}=R\,\sin^2\theta\,h_{yy}$, etc. Keeping the terms that are dominant in the limit $R\to\infty$, we obtain
\begin{eqnarray}
\Sigma_{xx}&=& \sigma+\frac{\sigma}{2}\p{h_y^2-h_x^2}+ \frac{\kappa}{2}
\p{h_{yy}^2-h_{xx}^2} \nonumber\\&+& \kappa h_x \partial_x \nabla^2 h \,,\\\Sigma_{yx}&=&-\sigma h_x h_y-\kappa h_{xy} \nabla^2 h+\kappa h_y  \partial_x \nabla^2 h,\\
\Sigma_{zx}&=&  \sigma h_x-\kappa \partial_x \nabla^2 h,\\
\Sigma_{yy}&=& \sigma+\frac{\sigma}{2}\p{h^2_{x}-h^2_{y}}+ \frac \kappa 2
\p{h^2_{xx}-h^2_{yy}} \nonumber\\&+& \kappa h_y \partial_y \nabla^2 h \,,\\
\Sigma_{xy}&=& - \sigma h_x h_y-\kappa h_{xy} \nabla^2 h+\kappa h_x  \partial_y \nabla^2 h,\\
\Sigma_{ry}&=&  \sigma h_y-\kappa  \partial_y \nabla^2 h,
\end{eqnarray} 
which agree with~\cite{Fournier07}.

\section{Correlation functions and relations among them} 
\label{Annexe_2}

There are five fundamental correlation functions, which we have calculated explicitly:
\begin{eqnarray}
\mbox{\textit{see eqs.~(\ref{upup}--\ref{ututpp}).}}
\nonumber
\end{eqnarray}
The other correlation functions either vanish or may be deduced from them, as explained below.
Using the relation:
\begin{eqnarray}
&&
\frac{\partial^{n+1}{\Ylm}(\theta,\phi)}{\partial\phi^{n+1}}
\frac{\partial^{p-1}{\Ylm}(\theta,\phi)^*}{\partial\phi^{p-1}}
= \nonumber\\
&&\qquad\qquad\qquad\qquad
-\frac{\partial^{n}{\Ylm}(\theta,\phi)}{\partial\phi^{n}}
\frac{\partial^{p}{\Ylm}(\theta,\phi)^*}{\partial\phi^{p}}\,,
\end{eqnarray}
which holds also in the presence of derivatives with respect to $\theta$, one deduces the following rule: when averaging a product of two terms one may pass
a derivative with respect to $\phi$ from one term to the other one while multiplying by
$-1$. hence,
\begin{eqnarray}
\langle u_{\phi}^2 \rangle &=& - \langle u
\, u_{\phi\phi}\rangle \, ,
\label{upup1}\\
\langle u_{\phi}\, u_{\theta} \rangle &=& - \langle u\, 
u_{\phi\theta}\rangle \, ,\\
\langle u_{\phi}\, u_{\theta\theta} \rangle &=& - \langle u_\theta\, 
u_{\phi\theta}\rangle \, .
\end{eqnarray}
An interesting consequence is that averages implying an odd number of derivatives
with respect to $\phi$ vanish: 
\begin{eqnarray}
\langle u\, u_\phi \rangle &=& 0\, ,\\
\langle u_\theta \, u_\phi \rangle &=& 0\, ,\\
\langle u_\phi \, u_{\phi\phi} \rangle &=& 0\, ,\\
\langle u_\theta \, u_{\theta\phi} \rangle &=& 0\, ,\\
\langle u_{\theta\theta} \, u_{\theta\phi} \rangle &=& 0\, ,\\
\langle u_{\phi\phi} \, u_{\theta\phi} \rangle &=& 0\, .
\label{upputp}
\end{eqnarray}
\begin{figure*}
\rule{18cm}{0.4pt}
\begin{eqnarray}
\average{\up^2}&=&\sum_{\omega,\omega'} \partial_\phi \Ylm(\theta,\phi)\,\partial_{\phi}
Y_{l'}^{m'}(\theta,\phi) \average{\ulm u_{l'm'}}=\sum_\omega\frac{k_B T}{\Hl}
\partial_\phi \Ylm \partial_\phi\!\p{\Ylm}^*=\sum_\omega  \frac{k_B T}{\Hl} m^2
|\Ylm|^2 \, ,\label{upup} \nonumber \\
&=&\sin^2\theta\,\frac{k_B T}{4 \pi} \sum_{l=2}^L
\frac{l \p{l+1}\p{2l+1}}{2 \Hl}\,,\\
\average{u_{\phi\phi}^2}&=&\sum_\omega  \frac{k_B T}{\Hl} m^4
|\Ylm|^2=\sin^2\theta\,\frac{k_B T}{4 \pi} \sum_{l=2}^L \frac{l \p{l+1}\p{2l+1}}{8
  \Hl}\pq{4+3\p{-2+l+l^2}\sin^2\theta}\, ,\\
\average{\ut^2}&=& \sum_\omega\frac{k_B T}{\Hl}\,\partial_\theta \Ylm
\partial_\theta\!\p{\Ylm}^*=\frac{k_B T}{4 \pi} \sum_{l=2}^L \frac{l
  \p{l+1}\p{2l+1}}{2 \Hl} \, ,\\
\average{u_{\theta \theta}^2}&=&\sum_\omega\frac{k_B T}{\Hl} \partial_\theta^2
\Ylm \,\partial_\theta^2\!\p{\Ylm}^*=\frac{k_B T}{4 \pi} \sum_{l=2}^L \frac{l
  \p{l+1}\p{2l+1}\p{-2+3 l +3 l^2}}{8 \Hl} \,, \\
\average{\ut u_{\theta \phi\phi}}&=&-\sum_\omega\frac{k_B T}{\Hl} m^2
\partial_\theta \Ylm \partial_\theta \p{\Ylm}^*=-\frac{k_B T}{4 \pi}
\sum_{l=2}^L\frac{l \p{l+1}\p{2l+1}}{2 \Hl} \pq{1+\frac{\p{l+3}\p{l-2}}{4} \sin^2\theta}
\, .
\label{ututpp}
\end{eqnarray} 
\rule{18cm}{0.4pt}
\end{figure*}
Starting from the addition theorem for spherical harmonics, one may show by suitable differentiations that the sum
\begin{equation}
\sum_{m=-l}^l \frac{\partial^k \Ylm(\theta,\phi)}{\partial \theta^k}
\frac{\partial^p {\Ylm}(\theta,\phi)^{*}}{\partial \theta^p}
\end{equation}
vanishes for $k+p$ odd, implying:
\begin{eqnarray}
\langle u \, u_\theta \rangle &=& 0 \, ,
\label{uut}\\
\langle u \, u_{\theta \theta \theta} \rangle &=& 0 \, .
\label{uuttt}
\end{eqnarray}
Note that this does not hold when derivations with respect to $\phi$
are also involved.

Starting again from the addition theorem, one may show that
\begin{eqnarray}
&&\sum_{m = -l}^l \frac{\partial^{k+1} \Ylm(\theta,\phi)}{\partial \theta^{k+1}}
\frac{\partial^{p-1} {\Ylm}(\theta,\phi)^{*}}{\partial \theta^{p-1}}
\nonumber\\
&&= -\!\sum_{m = -l}^l \frac{\partial^{k} \Ylm(\theta,\phi)}{\partial \theta^{k}}
\frac{\partial^{p} {\Ylm}(\theta,\phi)^{*}}{\partial \theta^{p}}\, .
\end{eqnarray}
It follows that, when averaging a product of two terms one may pass
a derivative with respect to $\theta$ from one term to the other one while multiplying by $-1$. This holds only, however, in the absence of derivatives with respect to $\phi$. As a consequence,
\begin{eqnarray}
\langle u_\theta \, u_\theta \rangle &=& - \langle u \, u_{\theta \theta}
\rangle \, ,\\
\langle u_\theta\, 
u_{\theta \theta \theta} \rangle  &=& - \langle u_{\theta \theta}
\, u_{\theta \theta } \rangle \, .
\end{eqnarray}

Finally, one may also use the fact that $\Delta \Ylm(\theta,\phi) = 0$, where $\Delta$
is the laplacian in spherical coordinates, to obtain
\begin{equation}
\langle u_\theta \, u_{\theta \theta } \rangle = \cot\theta \langle u
\, u_{\theta \theta } \rangle - \csc^2\theta \langle u_{\theta} \, 
u_{\phi \phi} \rangle \,.
\end{equation}
Since $\langle u_\theta \, u_{\theta \theta } \rangle$ vanishes, on obtains
\begin{equation}
\langle u_{\theta} \, u_{\phi \phi} \rangle = - \sin\theta \cos\theta \, \langle
u_\theta^2\rangle \, .
\label{utupp}
\end{equation}
In conclusion, eqs.~(\ref{upup}--{\ref{ututpp}), together
with eqs.~(\ref{upup1}--\ref{upputp}), eqs.~(\ref{uut}--\ref{uuttt}) and
eq.~(\ref{utupp}) give all the correlation functions of the derivatives of $u$.



\end{document}